\documentstyle[12pt,graphicx,pstricks,braket,amsmath,amssymb,enumitem,
                    epstopdf,pdfpages,cite,tabu,hyperref,caption]{article}
\numberwithin{equation}{section}
\begin{document}

\def\AEF{A.E. Faraggi}

\def\JHEP#1#2#3{{JHEP} {\textbf {#1}}, (#2) #3}
\def\vol#1#2#3{{\bf {#1}} ({#2}) {#3}}
\def\NPB#1#2#3{{\it Nucl.\ Phys.}\/ {\bf B#1} (#2) #3}
\def\PLB#1#2#3{{\it Phys.\ Lett.}\/ {\bf B#1} (#2) #3}
\def\PRD#1#2#3{{\it Phys.\ Rev.}\/ {\bf D#1} (#2) #3}
\def\PRL#1#2#3{{\it Phys.\ Rev.\ Lett.}\/ {\bf #1} (#2) #3}
\def\PRT#1#2#3{{\it Phys.\ Rep.}\/ {\bf#1} (#2) #3}
\def\MODA#1#2#3{{\it Mod.\ Phys.\ Lett.}\/ {\bf A#1} (#2) #3}
\def\RMP#1#2#3{{\it Rev.\ Mod.\ Phys.}\/ {\bf #1} (#2) #3}
\def\IJMP#1#2#3{{\it Int.\ J.\ Mod.\ Phys.}\/ {\bf A#1} (#2) #3}
\def\nuvc#1#2#3{{\it Nuovo Cimento}\/ {\bf #1A} (#2) #3}
\def\RPP#1#2#3{{\it Rept.\ Prog.\ Phys.}\/ {\bf #1} (#2) #3}
\def\APJ#1#2#3{{\it Astrophys.\ J.}\/ {\bf #1} (#2) #3}
\def\APP#1#2#3{{\it Astropart.\ Phys.}\/ {\bf #1} (#2) #3}
\def\EJP#1#2#3{{\it Eur.\ Phys.\ Jour.}\/ {\bf C#1} (#2) #3}
\def\etal{{\it et al\/}}
\def\notE6{{$SO(10)\times U(1)_{\zeta}\not\subset E_6$}}
\def\E6{{$SO(10)\times U(1)_{\zeta}\subset E_6$}}
\def\highgg{{$SU(3)_C\times SU(2)_L \times SU(2)_R \times U(1)_C \times U(1)_{\zeta}$}}
\def\highSO10{{$SU(3)_C\times SU(2)_L \times SU(2)_R \times U(1)_C$}}
\def\lowgg{{$SU(3)_C\times SU(2)_L \times U(1)_Y \times U(1)_{Z^\prime}$}}
\def\SMgg{{$SU(3)_C\times SU(2)_L \times U(1)_Y$}}
\def\Uzprime{{$U(1)_{Z^\prime}$}}
\def\Uzeta{{$U(1)_{\zeta}$}}

\newcommand{\cc}[2]{c{#1\atopwithdelims[]#2}}
\newcommand{\bev}{\begin{verbatim}}
\newcommand{\beq}{\begin{equation}}

\newcommand{\beqa}{\begin{eqnarray}}
\newcommand{\beqn}{\begin{eqnarray}}
\newcommand{\eeqn}{\end{eqnarray}}
\newcommand{\eeqa}{\end{eqnarray}}
\newcommand{\eeq}{\end{equation}}
\newcommand{\beqt}{\begin{equation*}}
\newcommand{\eeqt}{\end{equation*}}
\newcommand{\Eev}{\end{verbatim}}
\newcommand{\bec}{\begin{center}}
\newcommand{\eec}{\end{center}}
\newcommand{\bes}{\begin{split}}
\newcommand{\ees}{\end{split}}
\def\ie{{\it i.e.~}}
\def\eg{{\it e.g.~}}
\def\half{{\textstyle{1\over 2}}}
\def\nicefrac#1#2{\hbox{${#1\over #2}$}}
\def\third{{\textstyle {1\over3}}}
\def\quarter{{\textstyle {1\over4}}}
\def\m{{\tt -}}
\def\mass{M_{l^+ l^-}}
\def\p{{\tt +}}

\def\slash#1{#1\hskip-6pt/\hskip6pt}
\def\slk{\slash{k}}
\def\GeV{\,{\rm GeV}}
\def\TeV{\,{\rm TeV}}
\def\y{\,{\rm y}}

\def\l{\langle}
\def\r{\rangle}
\def\LRS{LRS  }

\begin{titlepage}
\samepage{
\setcounter{page}{1}
\rightline{LTH--998} 
\vspace{1.5cm}

\begin{center}
 {\Large \bf 
 Light $Z^\prime$ in Heterotic String Standard--like Models}
\end{center}

\begin{center}

P. Athanasopoulos\footnote{
		                  E-mail address: panos@liv.ac.uk}, 
A.E. Faraggi\footnote{
		                  E-mail address: alon.faraggi@liv.ac.uk}
and 
V.M. Mehta\footnote{ E-mail address:
	                          Viraf.Mehta@liv.ac.uk}
\\
\vspace{.25cm}
{\it Department of Mathematical Sciences\\
University of Liverpool, Liverpool, L69 7ZL, United Kingdom}
\end{center}

\begin{abstract}

The discovery of the Higgs boson at the LHC supports the 
hypothesis that the Standard Model provides an effective 
parameterisation of all subatomic experimental data up to
the Planck scale. String theory, which provides a viable
perturbative approach to quantum gravity, requires for its 
consistency the existence of additional gauge
symmetries beyond the Standard Model. 
The construction of heterotic--string models with a viable 
light $Z^\prime$ is, however, highly constrained. 
We outline the construction of standard--like heterotic--string
models that allow for an additional Abelian gauge symmetry
that may remain unbroken down to low scales.
We present a string inspired model, consistent with the
string constraints.

\end{abstract}
\smallskip}
\end{titlepage}

\section{Introduction}

String theory provides a predictive framework for exploring 
unification of the gravitational and gauge interactions.
The consistency of string theory dictates that it 
must accommodate a specific number of degrees of 
freedom to produce an anomaly free and finite 
framework. Some of these degrees of freedom
give rise to the gauge symmetries that we 
may identify with those of the subatomic 
gauge interactions, whereas the others do not
produce an observable manifestation in 
contemporary experiments. This is both a theoretical 
challenge, as well as a technological one,
since the hierarchy of the gravitational and gauge 
interactions implies that the additional degrees 
of freedom required by string theory are interacting
extremely weakly with its observable segments. 

The methodology to explore the string unification of 
gravity and the gauge interactions entails the 
construction of string models that reproduce the 
observed subatomic matter and interactions. Indeed,
numerous quasi--realistic models have been constructed
by using target--space and worldsheet techniques
\cite{spreview}. To date all these models possess $N=1$ spacetime
supersymmetry, which stabilises the constructions and
provides a better fit to the experimental data in
some scenarios. However, the question of supersymmetry
breaking is an open issue and it may well be that 
it is not manifested within reach of contemporary
experiments. The main problem in that case will be
to construct viable string models in which supersymmetry
is broken at a higher scale, which is not outside the realm of 
possibilities. The subatomic data is encoded in the 
Standard Model of particle physics, and therefore the 
realistic string constructions aim to reproduce
the Minimal Supersymmetric Standard Model. 
The Standard Model data provide hints that the 
matter states and gauge bosons originate from 
representations of larger symmetry groups. 
Most appealing in this context is the embedding of the 
Standard Model matter states in the three $\bf{16}$ spinorial 
representations of an $SO(10)$ gauge group. This 
structure is reproduced perturbatively in the heterotic--string.

The gauge content of the Standard Model consists of 
the three group sectors that correspond to the 
strong, electroweak and weak--hypercharge interactions. 
These correspond to a rank four group, whereas the 
heterotic--string in four dimensions may give rise 
to a rank--22 group. While the Standard Model states
in heterotic--string models are typically neutral 
under eight of these degrees of freedom, they 
are charged with respect to the others. The possible
observation of an additional gauge degree of freedom
at contemporary experiments will provide 
evidence for the additional degrees of freedom 
predicted by string theory. 

The existence of additional $U(1)$ gauge symmetries 
in string theory 
has indeed been of interest since the observation that
string theory is free of gauge and gravitational anomalies
\cite{greenschwarz}.
Indeed, extra $Z^\prime$ string inspired models occupy
a substantial number of studies that utilise effective 
field theory constructions to explore their phenomenological
implications \cite{zphistory, zpbminusl, zpfff,fm1}.  
However, quite surprisingly, the construction of 
quasi--realistic worldsheet heterotic--string models that 
accommodate an extra $U(1)$ gauge symmetry 
in the observable sector, which may remain 
unbroken at low scales, has proven to be an arduous 
task for a variety of phenomenological constraints. 
In fact, to date there does not exist a single
quasi--realistic exact string solution that 
accommodates an extra $U(1)$ gauge symmetry
that remains viable down to low scales. 
The problem stems from the fact that in many string
constructions the extra family universal $U(1)$s, that are 
typically discussed in string inspired models, are anomalous,
and cannot remain unbroken down to low scales.

On the other hand, models that give rise to anomaly free 
family universal extra $U(1)$ symmetries cannot accommodate the 
low scale gauge coupling data \cite{zpgcu}. The primary reason 
is that the charge assignment of the Standard Model
states under these anomaly free $U(1)$s does not admit an $E_6$ 
embedding, which emerges as a necessary ingredient to accommodate
the gauge coupling data. 
In \cite{zpgcu} we discussed the worldsheet construction 
of extra anomaly free $Z^\prime$ models that do admit an $E_6$ 
embedding. The observable gauge symmetry at the string level 
in the model of \cite{zpgcu} is $SO(6)\times SO(4)\times U(1)$, 
which is broken to $SU(3)_C\times SU(2)_W\times U(1)_Y\times U(1)_{Z^\prime}$, 
with the $U(1)_{Z^\prime}$ being anomaly free and admits an $E_6$ embedding.

In this paper we discuss the worldsheet construction in 
standard--like models, \ie in which the
observable gauge symmetry is broken at the string level to
$SU(3)_C\times SU(2)_W\times U(1)_{B-L}\times U(1)_{T_{3_R}}\times 
U(1)_{\zeta}$.
In both of these cases the symmetry is broken to 
$SU(3)_C\times SU(2)_W\times U(1)_{Y}\times U(1)_{Z^\prime}$,
by the Vacuum Expectation Value (VEV) 
of the Standard Model singlet in the $\bf{16}$ representation
of $SO(10)$. We use the tools of the free fermionic formulation 
for our analysis, which for the gauge degrees of the heterotic--string
is entirely equivalent to a free bosonic description \cite{heterotic}.

\section{Additional $U(1)$s in heterotic--string
models}\label{au1}

The heterotic--string models in the free fermionic formulation \cite{fff} produce some of the most realistic string models constructed to date. The quasi--realistic models correspond to $Z_2\times Z_2$ orbifold compactifications, at special points in the moduli space, with discrete Wilson lines \cite{z2xz2}. They lead to a rich space of three generation models charged under a subgroup of $SO(10)$. In the free fermionic formulation all the degrees of freedom needed to cancel the conformal worldsheet anomaly are represented in terms of free fermions propagating on the string worldsheet. For example, a set of eight complex fermions give rise to the Cartan generators of the observable gauge group and are denoted by $\left\{\overline{\psi}^{1,\cdots,5}, \overline{\eta}^{1,2,3}\right\}$. Under parallel transport around the non-contractible loops of the worldsheet torus, these fermions pick up a phase. The phases of the worldsheet fermions, constrained by modular invariance, then make up our boundary condition basis vectors which, in addition to the associated one--loop GGSO coefficients, describe the heterotic--string models in the free fermionic formulation fully \cite{fff}. 
 
The basis vectors span a finite additive group, $\Xi$, consisting of the sectors, $\alpha$, from which the physical states are obtained by acting on the vacuum with bosonic and fermionic oscillators and by applying the GGSO projections.

For a sector consisting of periodic complex fermions only, the vacuum is a spinor, $\ket{\pm}$, representing the Clifford algebra of the corresponding zero modes, $f_0$ and $f^*_0$, which have fermion number $F(f)=0,-1$ respectively.  In addition, the Cartan subalgebra of our rank--22 group is $U(1)^{22}$, generated by the right--moving currents, $\overline{f}\overline{f}^*$.  For each complex fermion, $f$, the $U(1)$ charges correspond to
\begin{align}
Q(f)=\frac 12\alpha(f)+F(f).
\label{Qf}
\end{align}
The representation (\ref{Qf}) shows that $Q(f)$ is identical to the worldsheet fermion numbers, $F(f)$, for worldsheet fermions with Neveu--Schwarz boundary conditions, $\alpha(f)=0$, and is $F(f)+\frac 12$ for those with Ramond boundary conditions, $\alpha(f)=1$. The charges for the $\vert\pm\rangle$ spinor vacua are $\pm\frac 12$.

The boundary conditions of the set of eight complex worldsheet 
fermions that give rise to the Cartan generators
of the observable gauge group, with ${\bar\psi}^{1,\cdots,5}$ generating the
$SO(10)$ group and ${\bar\eta}^{1,2,3}$
generating three $U(1)$ symmetries, denoted by $U(1)_{1,2,3}$, will be the focus of our discussion in this paper.
The vector bosons contributing to 
the four dimensional observable gauge group are charged with respect 
to these Cartan generators, and arise from the untwisted sector, 
as well as from twisted sectors, \ie sectors that contain
periodic fermions.

The early three generation free fermionic models were NAHE based 
models \cite{nahe} with more recent methods for the systematic 
classification of free fermionic models developed in 
\cite{typeIIclass,classification,bfgrs}. 
In NAHE based models \cite{fsu5,slm,alr,lrs,su421} 
the first set of five basis vectors,
$\{{{\bf 1},S,b_1,b_2,b_3}\}$, is fixed.  The addition of $b_1$, $b_2$ and $b_3$ breaks the $N=4$ spacetime SUSY, generated by $S$, to $N=1$ and the respective sectors correspond to the 
three twisted sectors of the $Z_2\times Z_2$ orbifold.
At this stage, the gauge symmetry is $SO(10)\times SO(6)^3\times E_8$ with the hidden $E_8$ being generated by $\left\{\bar{\phi}^{1,\dots,8}\right\}$. 
Adding the basis vector 
$x\equiv\{{\bar\psi}^{1,\cdots,5}, {\bar\eta}^{1,2,3}\}\equiv 1,$
produces the extended NAHE basis set \cite{enahe} with the resulting gauge symmetry being $E_6\times U(1)^2\times SO(4)^3\times E_8$, 
where the linear combination 
$J_\zeta={\bar\eta}^{1\ast}{\bar\eta}^1+
        {\bar\eta}^{2\ast}{\bar\eta}^2+
        {\bar\eta}^{3\ast}{\bar\eta}^3$
generates the $U(1)$ charges in the decomposition of 
$E_6\rightarrow SO(10)\times U(1)$. As we discuss below 
the vector $x$ plays a crucial role in generating a viable
light $Z^\prime$ in free fermionic models.

The next stage in constructing NAHE--based models involves
adding basis vectors to the NAHE set.
These additional vectors reduce the number of chiral generations 
to three and simultaneously break the four dimensional 
gauge group. The visible $SO(10)$ gauge symmetry is broken to one
of its maximal subgroups:
\begin{enumerate}[label=\Roman*]
\item 
\begin{enumerate}[label=\roman*]
\item $SU(5)\times U(1)$ (FSU5) \cite{fsu5}; 
\item $SU(3)\times SU(2)\times U(1)^2$ (SLM) \cite{slm}; 
\item $SO(6)\times SO(4)$ (PS) \cite{alr}; 
\end{enumerate}
\item 
\begin{enumerate}[label=\roman*]
\item $SU(3)\times U(1)\times SU(2)^2$ (LRS) \cite{lrs};
\item $SU(4)\times SU(2)\times U(1)$ (SU421) \cite{su421}. 
\end{enumerate}
\end{enumerate}

The difference between the models in case I and those in case II is 
the anomalous $U(1)_A$ symmetry that arises \cite{cleaverau1}.
In case I, the $U(1)_{1,2,3}$, as well as their linear combination
\begin{align}
U(1)_\zeta = U(1)_1+U(1)_2+U(1)_3,
\label{u1zeta}
\end{align}
are anomalous, whereas in the models of case II they are anomaly free.
This can be seen from the different symmetry breaking patterns: In the first case, $E_8\times E_8$, generated by the basis set 
$\left\{{\bf1},{S},\zeta,{x}\right\}$, breaks to $SO(16)\times SO(16)$ due to the choice of GGSO phases.  Implementing $b_1$ and $b_2$ then breaks $SO(16)\times SO(16)\to SO(10)\times U(1)^3\times SO(16)$.  We may also achieve this by breaking $E_8\times E_8\to E_6\times U(1)^2\times E_8$ via the addition of $b_1$ and $b_2$ as an initial step.  The gauge symmetry is then reduced to $SO(10)\times U(1)_\zeta\times U(1)^2\times SO(16)$ by GGSO projections that are equivalent to Wilson line breaking (\eg \cite{kk}).
The $U(1)_\zeta$ becomes anomalous because of the $E_6$ breaking to $SO(10)\times U(1)_\zeta$ and the 
GGSO projections removing states that would populate the $\mathbf{27}$ representation \cite{cleaverau1}. 
On the other hand, the models in case II are constructed from vacua with an $E_7\times E_7$ gauge symmetry.
These models circumvent the $E_6$ embedding, hence
$U(1)_\zeta$ may remain anomaly free. Only having the MSSM states 
survive to low scales produces an $SU(2)^2\times U(1)_\zeta$
mixed anomaly, which necessitates the existence of additional doublets
in the spectrum \cite{fm1}.
However, the charges of the additional doublets 
not possessing the $E_6$ embedding 
leads to disagreement with the experimental
gauge coupling data at the electroweak scale \cite{zpgcu}.
By contrast, if the charges do admit an $E_6$ embedding, 
the well known cancellation between the additional doublets and triplets
in the RGE solutions $\sin^2\theta_W(M_Z)$ and $\alpha_s(M_Z)$ \cite{zpgcu}, 
facilitates the compatibility with the gauge coupling data \cite{zpgcu}.   
We note that in both cases the relevant combination is
the identical combination of worldsheet currents given by 
$U(1)_\zeta$ in (\ref{u1zeta}). 

We remark that the string models produce several additional
$U(1)$s in the observable sector that may {\it a priori} give rise 
to a low scale $Z^\prime$. Two of those are the two combinations 
of $U(1)_{1,2,3}$, which are orthogonal to $U(1)_\zeta$. However, 
these are, in general, family non--universal and/or anomalous 
in the string models. 
Additionally, the models contain the 
combination\footnote{$U(1)_C=3/2U(1)_{B-L}$ and
$U(1)_L=2 U(1)_{T_{3_R}}$ are used in free fermionic models.},
$U(1)_C-U(1)_L$,
which is embedded in $SO(10)$, and is orthogonal 
to the weak--hypercharge \cite{zpbminusl}.
Here $Q_C$ and $Q_L$ are given in terms of the 
worldsheet charges by
\beq
Q_C=Q({\bar\psi}^1)+Q({\bar\psi}^2)+ Q({\bar\psi}^3)~
{\rm and } ~
Q_L=Q({\bar\psi}^4)+Q({\bar\psi}^5).
\label{qcql}
\eeq
However, this $U(1)$ combination has to be broken at a high scale 
to produce sufficient suppression of $m_{\nu_\tau}$. The reason
being the underlying $SO(10)$ symmetry at the string level, which
dictates that the $\tau$--neutrino Yukawa coupling is equal to that of 
the top quark. Hence, to produce a sufficiently suppressed 
mass term for $\nu_\tau$ requires a relatively high seesaw scale,
which is induced by the VEV of the Standard Model singlet in the 
$\bf{16}$ representation of $SO(10)$ \cite{tauneutrinomass}.

A light $Z^\prime$ in heterotic--string models must therefore
be a linear combination of $U(1)_C$, $U(1)_L$ and $U(1)_\zeta$. 
Thus, the $U(1)_\zeta$ symmetry, given by (\ref{u1zeta}), must
be anomaly free. Furthermore, the gauge coupling data dictate that 
the charges of the light states must admit an $E_6$ embedding. 
The task then is to obtain an anomaly free $U(1)_\zeta$, which admits
an $E_6$ embedding of the charges. However, as we noted above, in the 
quasi--realistic NAHE--based free fermionic
models \cite{fsu5,slm,alr,lrs,su421}, 
$U(1)_\zeta$ is either anomalous, or does not admit an 
$E_6$ embedding. 

We look for potential candidates in the space of symmetric orbifolds classified in \cite{classification}.  These models, generically, admit an anomalous $U(1)_\zeta$ due to its $E_6$ embedding.  However, a subset of these models may allow for an anomaly free $U(1)_\zeta$; the self--dual models under the spinor--vector duality of \cite{spinvecdual}.  The spinor--vector duality exchanges vectorial $\mathbf{10}$ representations of $SO(10)$ with spinorial $\mathbf{16}$ representations in the twisted sectors.  The self--dual models are those with an equal number of spinorial and vectorial representations.  $E_6$ is broken when these states arise from different twisted sectors.  A self--dual, three generation model with unbroken $SO(10)$ symmetry was presented in \cite{classification}, whereas such a model with a broken $SO(10)$ symmetry has not yet been constructed.

Another way to construct potential candidate models with an anomaly free $U(1)_\zeta$ is by following an alternative symmetry breaking pattern to $E_6\to SO(10)\times U(1)_\zeta$.  Previously this was accomplished by projecting out the enhancing gauge bosons originating in the $x$--sector, \ie those transforming in the $\mathbf{128}$ of $SO(16)$ that enhance $SO(16)\to E_8$.  Here we may build models that keep these enhancing gauge bosons but project out some of the $SO(10)$ gauge bosons. This will break $E_6$ to a different subgroup, as shown, for example, in the three generation $SU(6)\times SU(2)$ models of \cite{bfgrs}. 
The Standard Model generations 
are then embedded in the $({\bf15},{\bf1})$ and $({\bf6},{\bf2})$ 
representations of $SU(6)\times SU(2)$, \ie
all the states in the ${\bf27}$ of $E_6$ are retained
in the spectrum. The recipe, therefore, for 
constructing heterotic--string models with anomaly free 
$U(1)_\zeta$ is to retain the states arising from the
$x$--basis vector. In this case the untwisted gauge symmetry is
enhanced by the spacetime vector bosons arising from $x$. 
At the same time the twisted matter states from a given sector
$\alpha\in\Xi$ are complemented by the states from the sector
$\alpha+x$ to form complete $E_6$ representations, decomposed 
under the unbroken gauge symmetry at the string scale. 

\section{Standard--like models with light $Z^\prime$}\label{slmzp}

In \cite{zpgcu} we discussed the 
construction of Pati--Salam heterotic--string models with 
an anomaly free $U(1)_{Z^\prime}$, along the lines outlined
at the end of section \ref{au1}. In this section we 
articulate the construction of Standard--like heterotic string 
models with an anomaly free $U(1)_{Z^\prime}$.
The low scale $Z^\prime$ in the string models is a combination 
of the Cartan generators, $U(1)_{1,2,3}$, that are generated by the 
right--moving complex worldsheet fermions ${\bar\eta}^{1,2,3}$,
together with a $U(1)$ symmetry, which is embedded in the $SO(10)$ 
and is orthogonal to the weak--hypercharge. 

The vector bosons that generate the four dimensional 
gauge group in the free fermionic models arise
from three sectors:
the untwisted sector; the sector
$x$; and the sector
$\zeta=1+b_1+b_2+b_3=\{{\bar\phi}^{1, \cdots, 8}\}\equiv1$. 
The basis set $\{1,S, x, \zeta\}$ 
results in a four dimensional model with $N=4$ spacetime supersymmetry.
This model will have, at a generic point in the compactified space,
either $E_8\times E_8$ or $SO(16)\times SO(16)$ gauge symmetry depending 
on the GGSO phase $c({x\atop\zeta})=\pm1$.
In the $E_8\times E_8$ case,
the generators of the observable $E_8$ originate in the untwisted and in the $x$--sector,
with the adjoint of $SO(16)$ coming from the untwisted sector and the enhancing gauge bosons, transforming in the $\mathbf{128}$, originating in the $x$--sector.

Spacetime supersymmetry is broken to $N=1$ by the addition of the basis vectors $b_1$ and $b_2$.
This also reduces the observable gauge symmetry from $E_8\rightarrow  E_6\times U(1)^2$ or
$SO(16)\rightarrow SO(10)\times U(1)^3$.  The gauge symmetry can be reduced even further by 
additional vectors. With the exception of the model
in \cite{bfgrs}, the quasi--realistic free fermionic 
models follow the second symmetry breaking pattern, {\it i.e.}  
the vector bosons arising from the 
$x$--sector are, in all these models, projected out.

We consider the symmetry breaking pattern in the observable
sector induced by the following boundary condition assignments
in three consecutive basis vectors:
\beqn
&1.&b\{
{{\bar\psi}^{1,\cdots5,},{\bar\eta}^{1,2,3}}
     \}
=\{1~ 1\, 1\, 0\, 0\, 1\, 1\, 1 \}\,
  \Rightarrow SO(6)\times SO(4)
\label{s064breakingbc}\\
&2.&b\{
{{\bar\psi}^{1,\cdots5,},{\bar\eta}^{1,2,3}}
     \}
=\{1~ 1\, 0\, 1\, 0\, 1\, 1\, 1 \}\,
  \Rightarrow SO(4)\times SO(2)\times SO(2)\times SO(2)
\label{s04111breakingbc}\\
&3.&b\{
{{\bar\psi}^{1,\cdots5,},{\bar\eta}^{1,2,3}}
     \}=
\{
{1\over2}{1\over2}{1\over2}{1\over2}{1\over2}{1\over2}{1\over2}{1\over2}
\}
\Rightarrow SU(2)\times U(1)\times U(1)\times U(1)\times U(1),~~~~~~~
\label{su51breakingbc}
\eeqn
where on the right--hand side we display the breaking
pattern of the untwisted $SO(10)$ generators, induced by 
the consecutive basis vectors, and we omitted
the common factor of $U(1)^3$ corresponding to ${\bar\eta}^{1,2,3}$.
We consider here only 
the models with symmetric boundary conditions
for the set of
real fermions $\{y, \omega|{\bar y}, {\bar\omega}\}^{1,\dots,6}$.
The boundary condition
assignments for ${\bar\eta}^{1,2,3}$ 
are fixed by the modular invariance constraints on
$N_{ij}(v_i\cdot b_j) = 0\ {\rm mod}\ 4$, whereas the modular invariance
constraints on the three additional basis vectors are fixed
by the boundary conditions of the worldsheet fermions
$\{{\bar\phi}^{1,\dots,8}\}$, which produce the Cartan
generators of the hidden sector gauge group. 

We denote the three vectors that extend the NAHE--set by $\alpha$,
$\beta$ and $\gamma$. Each of these vectors then incorporates 
one of the boundary condition assignments given in (\ref{s064breakingbc}),
(\ref{s04111breakingbc}) and 
(\ref{su51breakingbc}), respectively. The vector $x$ may then
arise as, for example, the vector $2\gamma$, or as a separate basis
vector. The requirement is, however, that the vector bosons arising 
from the $x$--sector are retained in the spectrum. 

\begin{figure}[ht]
\begin{center}
\includegraphics[scale=1]{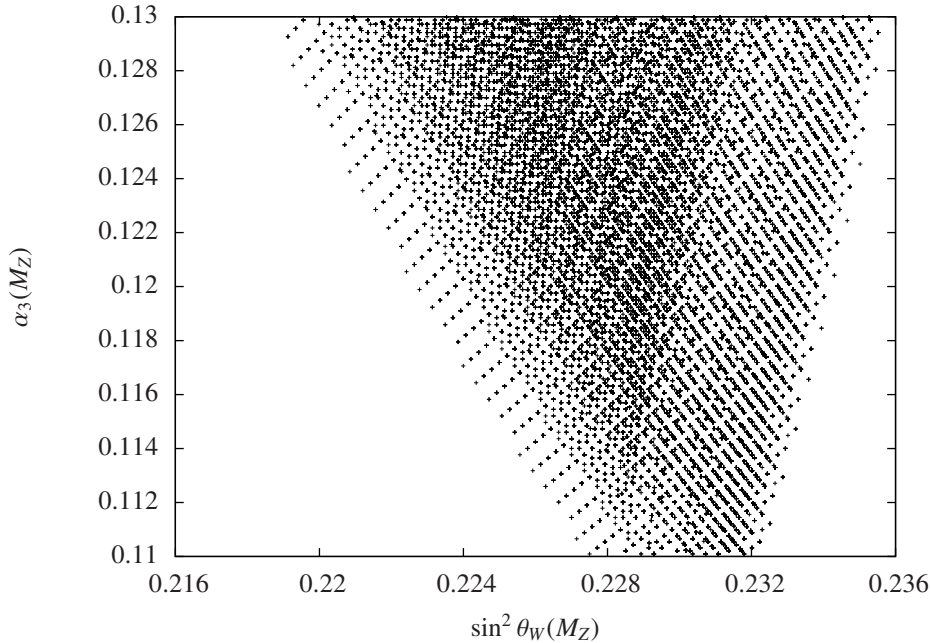} 
\caption{\it Solutions of the gauge coupling RGEs in the presence of an additional $U(1)$ symmetry with $E_6$ embedding \cite{VMMThesis}, assuming string unification between \\ $2\cdot 10^{16}\leq\mu\leq 5.27\cdot 10^{17}\,\mathrm{GeV}$ and $0< \alpha_{\mathrm{string}}\leq 0.1$. The phenomenologically viable region corresponds to $M_{\mathrm{string}}\sim 2\cdot 10^{16}\,\mathrm{GeV}$ within this range of $\alpha_{\mathrm{string}}$.  
}
\label{figure1}
\end{center}
\end{figure}

The untwisted gauge symmetry arising from the untwisted 
vector bosons after implementation of the GGSO projections 
of the basis vectors $\alpha$, $\beta$ and $\gamma$ is 
\beq
SU(2)\times U(1)_C\times 
U(1)_{{\bar\psi}^3}\times 
U(1)_{{\bar\psi}^4}\times 
U(1)_{{\bar\psi}^5}\times 
U(1)_{{\bar\eta}^1}\times 
U(1)_{{\bar\eta}^2}\times 
U(1)_{{\bar\eta}^3},
\label{4dggabc}
\eeq
where 
$Q_C=Q({\bar\psi}^1)+Q({\bar\psi}^2)$ and 
we denoted in (\ref{4dggabc}) the
worldsheet fermions that generate each $U(1)$ symmetry. 
The inclusion of the spacetime vector bosons that survive
the GGSO projections from the $x$--sector
then enhances the untwisted gauge symmetry to
\beq
SU(3)_C\times 
SU(2)_L\times 
U(1)_{C^\prime}\times 
U(1)_{4^\prime}\times 
U(1)_{5^\prime}\times
U(1)_{1^{\prime\prime}}\times 
U(1)_{2^{\prime\prime}},
\label{xenhanced}
\eeq
where
\beqn
U(1)_{3^\prime} & = & U(1)_{{\bar\psi}^3} 
                    + U(1)_{{\bar\psi}^4} 
                    + U(1)_{{\bar\psi}^5}
                    - U(1)_\zeta; \label{u13}\\
U(1)_{2^\prime} & = & U(1)_C + U(1)_{{\bar\psi}^3} 
                    + U(1)_{{\bar\psi}^4} 
                    + U(1)_{{\bar\psi}^5}
                    + U(1)_\zeta; \label{u12}\\
U(1)_{C^\prime} 
      & = &3 U(1)_C - U(1)_{{\bar\psi}^3} 
                    - U(1)_{{\bar\psi}^4} 
                    - U(1)_{{\bar\psi}^5}
                    - U(1)_\zeta; \label{u1cp}\\
U(1)_{4^\prime}       
      & = &           U(1)_{{\bar\psi}^4} 
                    - U(1)_{{\bar\psi}^5}; \label{u14p}\\
U(1)_{5^\prime} 
      & = &         2 U(1)_{{\bar\psi}^3}  
                    - U(1)_{{\bar\psi}^4} 
                    - U(1)_{{\bar\psi}^5}; \label{u15p}\\
U(1)_{1^{\prime\prime}} 
      & = &           U(1)_{{\bar\eta}^1}
                    - U(1)_{{\bar\eta}^2}; \label{u11p}\\
U(1)_{2^{\prime\prime}}
      & = &           U(1)_{{\bar\eta}^1}
                    + U(1)_{{\bar\eta}^2}
                    - 2 U(1)_{{\bar\eta}^3}.\label{u12p}
\eeqn 
$U(1)_{3^\prime}$ and $U(1)_{2^\prime}$ are the combinations that
are embedded in $SU(3)_C$ and $SU(2)_L$, respectively, and
$U(1)_\zeta$ is given by (\ref{u1zeta}).
The observable matter representations in the free fermionic models
arise from the sectors $b_j$, which produce states 
in the spinorial $\bf{16}$ representation of $SO(10)$, 
decomposed under the unbroken untwisted gauge group, 
and the sectors $b_j+x$, which produce states 
in the ${\bf10}+{\bf1}$ representations of $SO(10)$ that are 
decomposed similarly. Under the rotation of the 
Cartan generators displayed in (\ref{u13}--\ref{u12p}),
the states from these sectors combine to form
representations of the enhanced gauge group
in (\ref{xenhanced}). We can make a
further rotation on the $U(1)$ generators by taking
\beqn
U(1)_{C^{\prime\prime}} 
      & = &{1\over 4} U(1)_{C^\prime} 
        -  {1\over 2} U(1)_{5^\prime}; \label{u1cpp}\\
U(1)_{\zeta^\prime}       
      & = &{1\over 4} U(1)_{C^\prime} 
        +  {1\over 2} U(1)_{5^\prime}. \label{u1zetap}
\eeqn 
This reproduces the charge assignments in the ${\bf27}$ representation
of $E_6$, which are displayed in Table \ref{table27rot}. 
Additionally, the model contains pairs of heavy Higgs 
states 
\beq
{\cal N}+{\bar{\cal N}}=  (1,1, {3\over2},-1, {1\over2})+
                          (1,1,-{3\over2},+1,-{1\over2})
\label{heavyhiggs}
\eeq
that are needed to break the gauge symmetry to 
$SU(3)_C\times SU(2)_L\times U(1)_Y\times U(1)_{Z^\prime}$, 
where the $U(1)_Y$ and $U(1)_{Z^\prime}$ combinations are given by
\beqn
U(1)_Y & = &      {1\over3} U(1)_{C^{\prime\prime}} 
                + {1\over2} U(1)_{4^\prime}~,\label{u1y}\\
U(1)_{Z^\prime} & = & 
                  {1\over3} U(1)_{C^{\prime\prime}} 
                - {1\over3} U(1)_{4^\prime}
                - {5\over3} U(1)_{\zeta^\prime}. 
\label{u1zprime}
\eeqn
The model also contains a pair of vector--like light Higgs states that
are needed to obtain agreement with the gauge coupling data
at the electroweak scale,
\beq
h + {\bar h} = (1,2, 0, -1, +1)+
               (1,2, 0, +1, -1). 
\label{lighthiggs}
\eeq
The vector--like nature of the additional electroweak doublet 
pair is required because of anomaly cancellation.
In Figure \ref{figure1} we demonstrate that this spectrum,
assuming unification of the couplings at the heterotic--string
scale, is in agreement with $\sin^2\theta_W(M_Z)$ and $\alpha_3(M_Z)$. 
\begin{table}[!h]
\noindent 
{\small
\begin{center}
{\tabulinesep=1.2mm
\begin{tabu}{|l|cc|c|c|c|}
\hline
Field &$\hphantom{\times}SU(3)_C$&$\times SU(2)_L $
&${U(1)}_{C^{\prime\prime}}$&${U(1)}_{4^\prime}$&${U(1)}_{\zeta^\prime}$\\
\hline
$Q_L^i$&    $3$       &  $2$ &  $+\frac{1}{2}$    &        
                               ~~$ 0$     &    ~~$\frac{1}{2}$    \\
$u_L^i$&    ${\bar3}$ &  $1$ &  $-\frac{1}{2}$    &        
                               $-1$     &    ~~$\frac{1}{2}$    \\
$d_L^i$&    ${\bar3}$ &  $1$ &  $-\frac{1}{2}$    &        
                               $+1$    &    ~~$\frac{1}{2}$    \\
$e_L^i$&    $1$       &  $1$ &  $+\frac{3}{2}$    &        
                               $+1$     &    ~~$\frac{1}{2}$    \\
$L_L^i$&    $1$       &  $2$ &  $-\frac{3}{2}$    &        
                               ~~$0$     &    ~~$\frac{1}{2}$    \\
$N_L^i$&    $1$       &  $1$ &  $+\frac{3}{2}$    &        
                               $-1$     &    ~~$\frac{1}{2}$    \\
\hline
$D^i$       & $3$       & $1$ &  $-1$  &   ~~$ 0$     &    $-1$    \\
${\bar D}^i$& ${\bar3}$ & $1$ &  $+1$  &   ~~$ 0$     &    $-1$    \\
${\bar H}^i$& $1$       & $2$ &  ~~$0$   &   $+1$     &    $-1$    \\
${H}^i$     & $1$       & $2$ &  ~~$0$   &   $-1$     &    $-1$    \\
\hline
$S^i$       & $1$       & $1$ &  ~~$0$  &   ~~$0$       &    $+2$    \\
\hline
\end{tabu}}
\end{center}
}
\caption{\label{table27rot}
\it 
High scale spectrum and
$SU(3)_C\times SU(2)_L\times U(1)_{C^{\prime\prime}}\times 
U(1)_{4^\prime}\times U(1)_{\zeta^\prime}$ 
quantum numbers, with $i=1,2,3$ for the three light 
generations. The charges are displayed in the 
normalisation used in free fermionic 
heterotic--string models. }
\end{table}

The superpotential of the model contains the following terms 
\beqn
&   &Q u {\bar H}+ Q d H + L e H + L N{\bar H} \\
& + &H{\bar H} S + D {\bar D} S \\
& + &QQ D+ ud {\bar D}+ dND + ue{D} +QL{\bar D} \\
& + &Qu{\bar h}+L N{\bar h} + h{\bar h}\phi,
\label{superpot}
\eeqn
where $\phi$ stands for generic $E_6$ singlet fields arising
in the string models and generation indices have been suppressed. 
The superpotential contains couplings of the electroweak doublets 
appearing in Table \ref{table27rot}, as well as of the additional
pair of electroweak doublets in (\ref{lighthiggs}). The identification of 
the electroweak Higgs doublets requires a detailed analysis 
of the renormalisation group evolution of the 
fermion and scalar couplings. Some of the couplings
appearing in (\ref{superpot}) should be suppressed 
by additional discrete symmetries \cite{discreet}
to ensure proton longevity.
Light neutrino masses are generated in the model by
the nonrenormalizable terms $NN{\bar{\cal N}}{\bar{\cal N}}$, 
which generate heavy Majorana masses for the right--handed 
neutrinos due to the VEV of the heavy Higgs states 
appearing in (\ref{heavyhiggs}). 
We note that the existence of an extra $Z^\prime$ 
at low scale necessitates the existence of the additional 
matter states at the low scale to guarantee that the 
spectrum is anomaly free. 

\section{Conclusions}

The Standard Model of particle physics provides a parameterisation for 
subatomic experimental data, which is in agreement with all 
observations to date. 
The recent discovery of the Higgs boson by the ATLAS and CMS experiments 
\cite{higgs}
at the LHC provides further evidence for the validity of the
Standard Model up to the Planck scale. Additional
support for this possibility stems from: matter gauge charges;
proton longevity; neutrinos mass suppression; logarithmic 
evolution of the Standard Model parameters in the gauge and
matter sectors. Preservation of the logarithmic running 
in the scalar sector of the Standard Model mandates its 
augmentation with a new symmetry, with supersymmetry being a
phenomenologically viable possibility. Ultimately,
we would like to calculate the parameters of the Standard Model
from a fundamental theory. String theory provides a consistent
framework to pursue this endeavour within a perturbatively 
finite theory of quantum gravity. 

A remarkable feature of string theory is that its consistency 
mandates the existence of additional gauge degrees of freedom.
Many of these extra degrees of freedom are expected to be broken
at a high scale or be hidden from the Standard Model states. 
Remarkably, however, while the construction of quasi--realistic
standard--like heterotic--string models has been achieved, 
the construction of such models with a light $Z^\prime$ 
has proven to be an arduous task.

In this paper we explored the construction of heterotic--string 
standard--like models with a viable $Z^\prime$
within the free fermionic formulation. The key 
in this construction is to maintain in the spectrum the
spacetime vector bosons from the $x$--sector that enhance 
the gauge symmetry arising from the untwisted sector. 
The result is that all the matter states from the ${\bf27}$ of 
$E_6$, decomposed under the final gauge group, 
are retained in the spectrum.
Concrete string models that realise this enhancement 
are the $SU(6)\times SU(2)$ heterotic--string models
of \cite{bfgrs}.
The outcome is that the family universal $U(1)_\zeta$ combination
in (\ref{u1zeta}) is anomaly free and agreement
with the gauge coupling data at the electroweak scale is 
facilitated. The search for heterotic--string standard--like 
models that realise this construction is currently underway. 

\section*{Acknowledgements}

AEF thanks Subir Sarkar and Theoretical Physics Department at the
University of Oxford for hospitality.
PA acknowledges support from the Hellenic State Scholarship
Foundation (IKY). 
This work was supported in part by the STFC (ST/G00062X/1).

\end{document}